\documentclass[twocolumn]{aastex62}
\usepackage{bm}
\received{March 7, 2018}
\revised{April 9, 2019}
\accepted{April 14, 2019}

\shorttitle{M87 black hole shadow}
\shortauthors{Kawashima, Kino \& Akiyama}

\begin{document}

\title{Black Hole Spin Signature in the Black Hole Shadow of M87 in  the Flaring State}

\correspondingauthor{Tomohisa Kawashima}
\email{kawashima.tomohisa@nao.ac.jp}

\author[0000-0001-8527-0496]{Tomohisa Kawashima}
\affil{National Astronomical Observatory of Japan, 
2-21-1 Osawa, Mitaka, Tokyo, 181-8588, Japan}

\author[0000-0002-2709-7338]{Motoki Kino}
\affil{National Astronomical Observatory of Japan, 
2-21-1 Osawa, Mitaka, Tokyo, 181-8588, Japan}
\affil{Kogakuin University of Technology \& Engineering,, Academic Support Center, 
2665-1 Nakano, Hachioji, Tokyo 192-0015, Japan}

\author[0000-0002-9475-4254]{Kazunori Akiyama}
\affil{National Astronomical Observatory of Japan, 
2-21-1 Osawa, Mitaka, Tokyo 181-8588, Japan}
\affil{MIT Haystack Observatory, 99 Millstone Road, Westford, MA 01886, USA}
\affil{National Radio Astronomy Observatory, 520 Edgemont Road, Charlottesville, VA 22903, USA}
\affil{Black Hole Initiative, Harvard University, 20 Garden Street, Cambridge, MA 02138, USA}








\newcommand{\kazu}[1]{\textcolor{cyan}{(Kazu: #1)}}
\newcommand{\kazuedt}[2]{\textcolor{cyan}{{#1} {\sout{#2}}}}
\newcommand{\kazuadd}[1]{\textcolor{cyan}{#1}}

\begin{abstract}
Imaging the immediate vicinity of supermassive black holes (SMBH) and 
extracting a BH-spin signature is one of the grand challenges in astrophysics.
M87 is known as one of the best targets for imaging the BH shadow and
it 
can be partially thick against synchrotron self-absorption (SSA), particularly in a flaring state with high mass accretion rate.
However, little is known about influences of the SSA-thick region on BH shadow images.
%
Here we investigate BH shadow images of M87 at 230~GHz 
properly taking into account the SSA-thick region.
When the BH has a high spin value, 
the corresponding BH shadow image shows
the positional offset between the center of the photon   ring and that of the SSA-thick ring at the innermost stable circular orbit 
(ISCO) due to the frame-dragging effect in the Kerr spacetime.
As a result, we find that a dark-crescent structure is generally produced 
 between  the photon ring  and the SSA-thick ISCO ring in the BH shadow image.
The scale size of the dark-crescent increase with BH spin: its width reaches up to $\sim 2$ gravitational radius when the BH spin is 99.8\% of its maximum value.
The dark crescent is regarded as a new signature of a highly spinning BH.
This feature is expected to appear in flaring states with relatively high mass accretion rate rather than the quiescent states.
We have simulated the image reconstruction of our theoretical image by assuming the current and future Event Horizon Telescope (EHT) array, and have found that the future EHT including space-very long baseline interferometry in 2020s can detect the dark crescent.
\end{abstract}



\section{Introduction} \label{sec:intro}

\begin{figure*}[t]
\begin{center}
\includegraphics[scale=0.57]{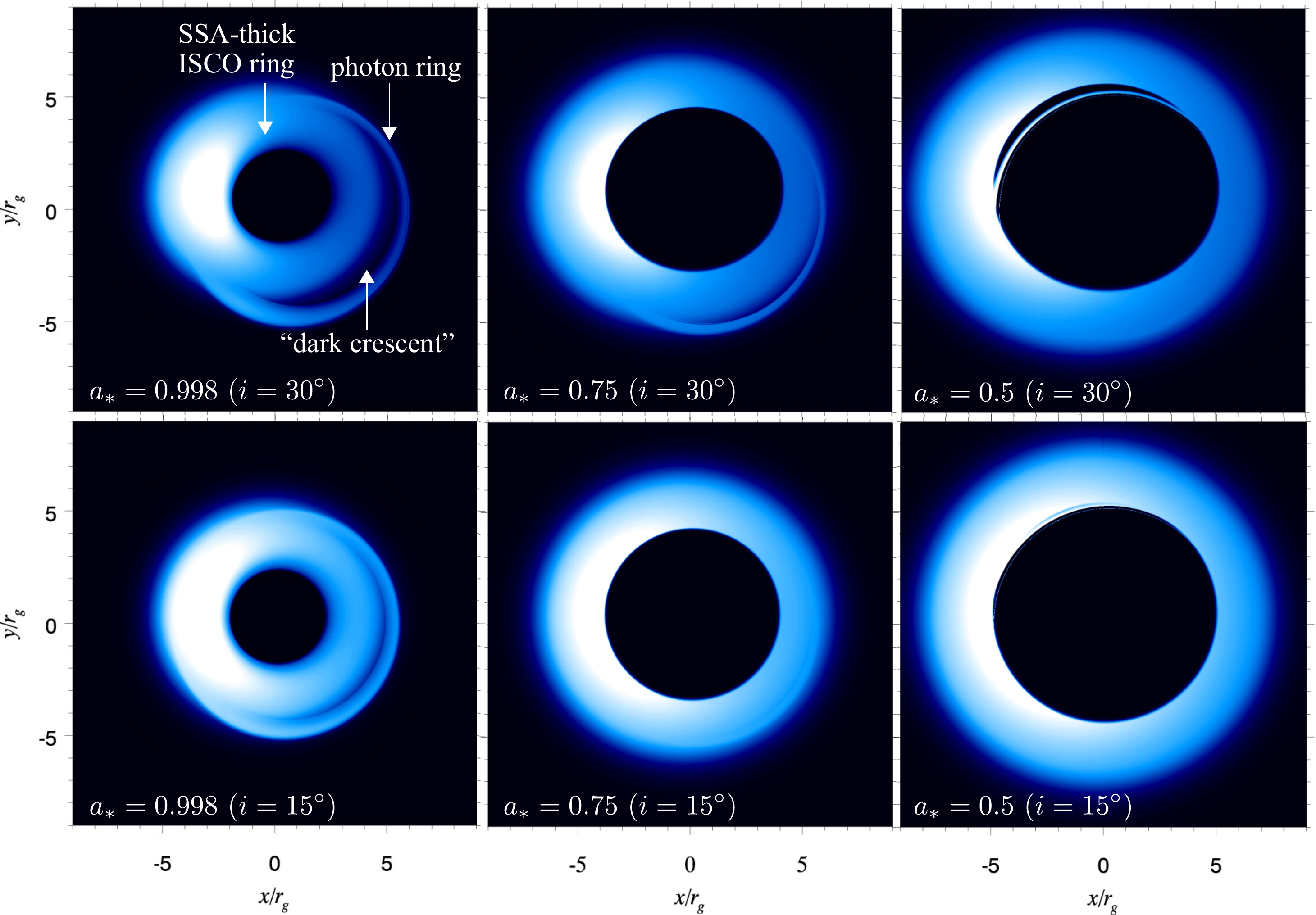}
\caption{Linear-scale intensity map of partially SSA-thick accretion flow onto a Kerr BH with $a_{*} = 0.998$ (left), 0.75 (center), and 0.5 (right), with viewing angle $i = 30^{\circ}$ (top) and ${15^{\circ}}$ (bottom) at $\nu = 230$ GHz. 
 The intensity is normalized by its maximum at each shadow image and the color bars are the same as those presented in Figure \ref{fig:shadow_tau}.
\label{fig:shadow_spin}
} 
\end{center}
\end{figure*}


 The Event Horizon 
 Telescope\footnote{\url{https://eventhorizontelescope.org/}} \citep[EHT;][]{2009astro2010S..68D} is a very long baseline
interferometry (VLBI) array aiming for imaging 
black hole (BH) shadows on event horizon scales 
and exploring fundamental  properties of BHs. 
The apparent size of the event
horizons of M87 at the center of the Virgo cluster 
and Sagittarius A* at the Galactic center 
are the two largest of all BH candidates.
Thus, these are the prime targets of the EHT, and
the early EHT observations have indeed
confirmed the existence of structures on the scale of the event horizons
 \citep{2008Natur.455...78D,2011ApJ...727L..36F,2012Sci...338..355D,2015ApJ...807..150A,2015Sci...350.1242J}.
The fully deployed EHT will have higher sensitivity and
greater dynamic range, which would enable EHT to image the "photon-ring", 
a surface brightness feature that appears on lines of sight passing close to the photon orbit
\citep[see][]{2015ApJ...814..115P,2016PhRvL.116c1101J}. 
The radius of the photon ring ($r_{\rm ph}$) is uniquely predicted by
general relativity \citep{1973blho.conf..241B,1979A&A....75..228L}.
Toward imaging BH shadows, 
new image reconstruction algorithms
to make accurate images from the EHT data have been also significantly 
developed and progressed recently
\citep[e.g.,][]{2016ApJ...829...11C,2016ApJ...820...90F,2017ApJ...838....1A,2017AJ....153..159A,2017ApJ...850..172J,2018ApJ...858...56K}. 
Quite recently, the EHT 2017 observations have shown the first image of BH shadow in M87 \citep{2019ApJ...875L...1E,2019ApJ...875L...2E,2019ApJ...875L...3E,2019ApJ...875L...4E, 2019ApJ...875L...5E, 2019ApJ...875L...6E}.
Quite recently, the EHT 2017 observations have shown the first image of BH shadow in M87 \citep{2019ApJ...875L...1E,2019ApJ...875L...2E,2019ApJ...875L...3E,2019ApJ...875L...4E,2019ApJ...875L...5E,2019ApJ...875L...6E}


One of the most fundamental quantities of a BH is
its angular momentum - commonly referred to as a spin.
Measuring the BH spin is one of the grand challenges in astrophysics.
The radius of the photon orbit strongly depends on BH spin.
Its radius changes from $3r_{\rm g}$ for a nonspinning BH to $r_{\rm g}$ for a maximally spinning BH for a prograde photon orbit on the equatorial plane in Boyer--Lindquist coordinates \citep{1972ApJ...178..347B}, where $r_{\rm g}\equiv GM_{\rm BH}/c^{2}$ is the gravitational radius of the BH with its mass $M_{\rm BH}$. 
However, the radius of the photon ring $r_{\rm ph}$ observed at infinity has  very weak dependence 
on the BH spin and viewing angle: the $r_{\rm ph}$ value changes only
less than $\sim$4\% from $5r_{\rm g}$, where
$r_{\rm ph}=\sqrt{27}r_{\rm g}$ for nonrotating BH and it has a very slightly decrease with BH spin \citep[see, e.g.,][and references therein]{2015ApJ...814..115P}.
The smallness of the change of $r_{\rm ph}$ makes it difficult to extract 
information about BH spin from BH shadow images.


To overcome this difficulty, here we revisit BH shadow images 
by focusing on the synchrotron self-absorption (SSA) effect onto  BH shadow images.
In previous theoretical works on the BH shadow in M87, the discussion was focused on the SSA-thin region even though SSA-thick part appears, or the models with SSA-thin plasma were studied intensively
 \citep{2009ApJ...697.1164B,2012MNRAS.421.1517D,2016A&A...586A..38M}.
According to \citet[][hearafter K15]{2015ApJ...803...30K}, however, 
the flux density of M87 detected by early EHT observations 
\citep{2012Sci...338..355D,2015ApJ...807..150A}
is  explained by the mixture of SSA-thin and SSA-thick emissions.
In addition, even if the plasma surrounding the BH is SSA thin in the quiescent state, SSA-thick emission will be realized in a flaring state with a relatively high mass accretion rate in M87.
In this paper, therefore, we calculate the BH shadow images, considering that the plasma is SSA thick in the vicinity of the BH, 
 and report a newly found  BH-spin signature in BH shadow images
referred to as a dark crescent. 
In this work, we adopt the BH mass and the distance to M87 from the Earth 
as $M = 6.2 \times 10^{9} ~M_{\odot}$\footnote{This BH mass estimate is rescaled to the adopted distance of $D=16.7$~Mpc from the value for $D=17.9$~Mpc reported in \citet{2011ApJ...729..119G}.} \citep{2009ApJ...700.1690G,2011ApJ...729..119G}
and $D=16.7 ~{\rm Mpc}$ 
\citep{2010A&A...524A..71B}, respectively. 



\section{GR Ray-tracing Radiative Transfer}

\begin{figure}[t]
\begin{center}
\includegraphics[scale=0.54]{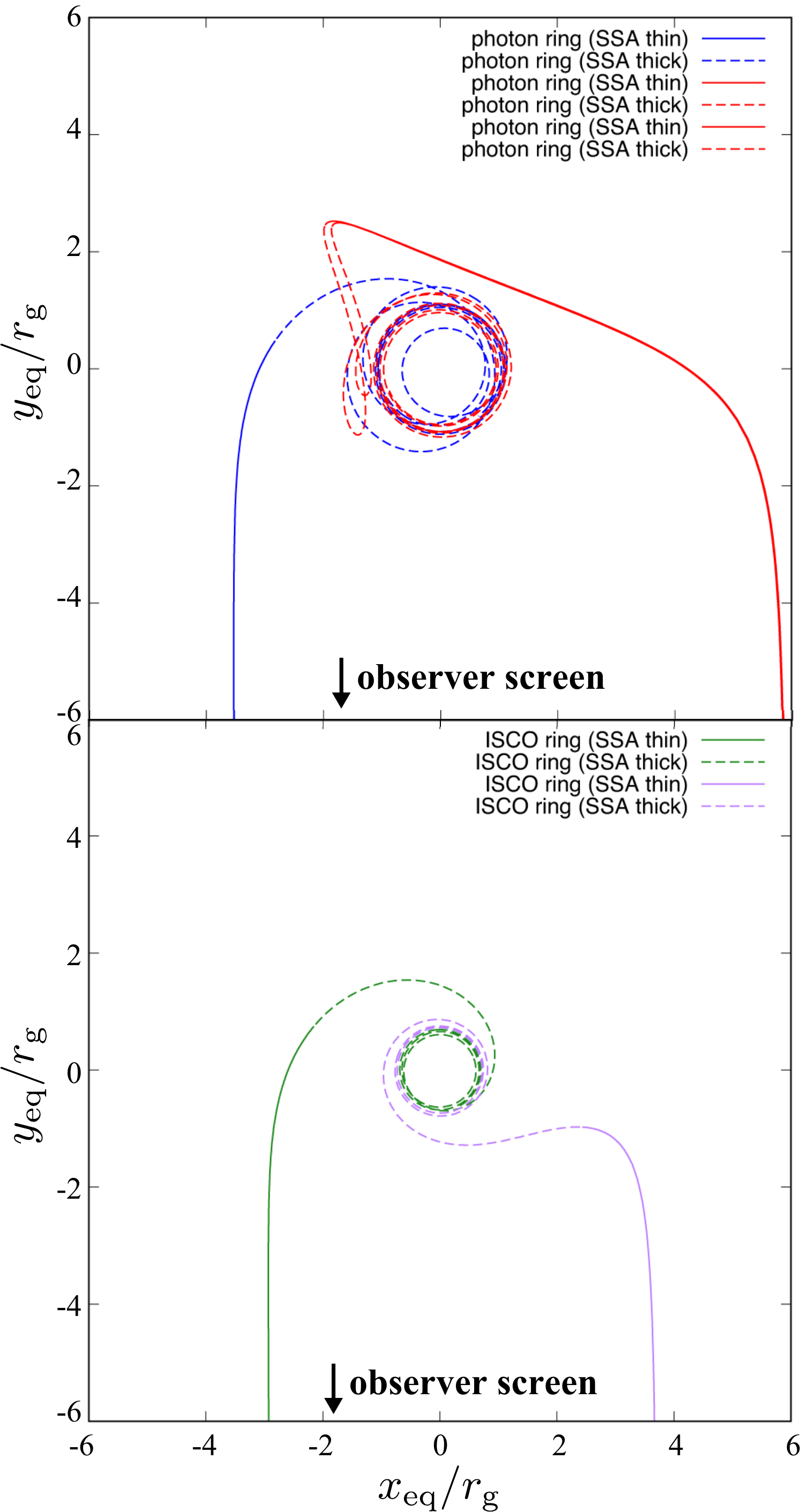}
\caption{Trajectories of photons forming the photon ring (top) and the SSA-thick ISCO ring (bottom), in the vicinity of the Kerr BH with $a_* = 0.998$. The viewing angle is $i=30^{\circ}$. Each trajectory is projected onto the equatorial plane of the Boyer--Lindquist coordinate (we note that we define $(x_{\rm eq}, y_{\rm eq})~ \equiv ~ (r \sin\theta \cos \varphi, r \sin \theta \sin \varphi)$). The solid/dashed curves represent the SSA-thin/SSA-thick regions along the rays. The positions of the rays on the observer screen are represented by the dashed lines in Figure \ref{fig:I_tau_1D}. }
\label{fig:trajectory}
\end{center}
\end{figure}

\begin{figure*}[ht!]
\begin{center}
\includegraphics[scale=0.58]{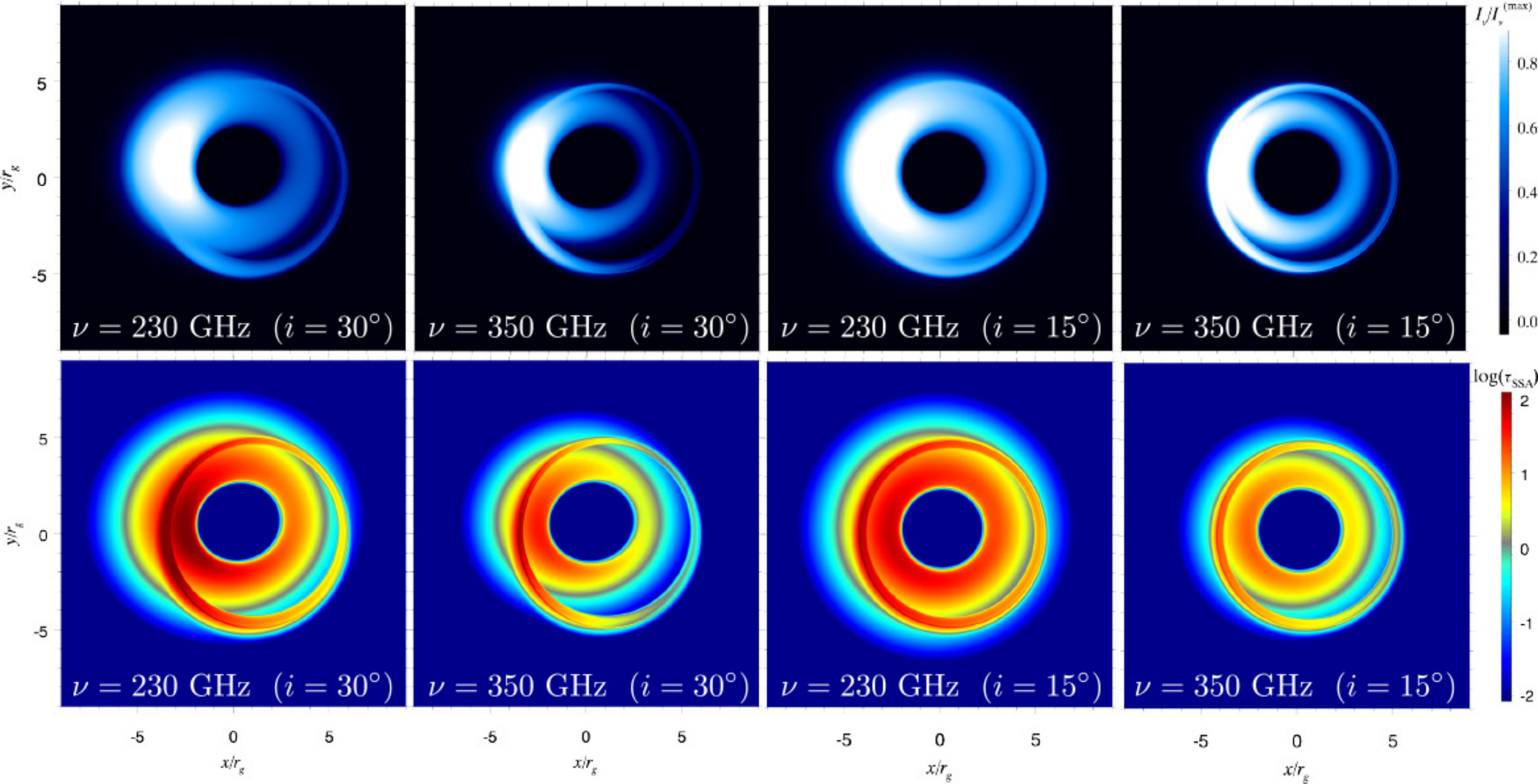}
\caption{Linear-scale intensity map (top) and log-scale $\tau_{\rm SSA}$ (bottom) in the observer's screen at 230~GHz (the first and the third columns) and 350~GHz (the second and the fourth columns). The viewing angle is $30^{\circ}$ (the two left columns) and $15^{\circ}$ (the right two columns). The BH spin is $a_* = 0.998$.}
\label{fig:shadow_tau}
\end{center}
\end{figure*}

Here, we briefly review our numerical code of GR ray-tracing radiative transfer.
We calculate the image of BH shadows by using our general relativistic ray-tracing 
radiative transfer code, \texttt{RAIKOU} (detail of the code are described in Kawashima et al. in preparation). 
We integrate the geodesic equations of photons by solving the Hamilton's canonical equations 
of motion describing time evolution of $r$, $\theta$, $\varphi$, $p_{r}$, $p_{\theta}$ 
of photons in the Boyer--Lindquist coordinate. Here, $r$, $\theta$, $\varphi$ are the radius, 
polar angle, and azimuthal angle, while $p_r$ and $p_\theta$ are the $r$ and $\theta$ components of covariant momentum vector \citep{2013ApJ...777...11S}. 
The ray-tracing solver is based on an eighth-order embedded Runge--Kutta method in which 
fifth- and third-order error estimators are used for the adoptive stepsize control 
\citep{2002nrca.book.....P}. 
As we investigate the radio emission from M87 in great detail, here we include the effects of
cyclo-synchrotron emission and absorption via the thermal electrons \citep{1996ApJ...465..327M}. 
The radiative transfer equations are integrated by using 
the observer-to-emitter method \citep{2012A&A...545A..13Y,2016ApJ...820..105P}. 
Our numerical code well reproduces photon trajectories 
\citep{1983grg1.conf....6C,2013ApJ...777...13C} and 
BH shadow images \citep{1979A&A....75..228L,2004ApJ...611..996T,2016ApJ...820..105P}
in the Kerr spacetime.

\section{Calculation setup for M87}

In this section, we describe the model parameter setup based on recent VLBI observations of M87. 
The spin of the BH of M87 is parameterized to be $a_{*} = $ 0.5, 0.75, and  0.998, 
as the BH spin of M87 is generally expected to be relatively high \citep{2012Sci...338..355D, 2018ApJ...868..146N, 2018ApJ...868...82T}. 
We set the viewing angle between the line of sight and the BH-spin vector to be $i = 30^{\circ}$ and $15^{\circ}$ in this paper, because
the viewing angle of M87 is constrained to be ${\lesssim} ~ 30^{\circ}$
by the recent 86~GHz VLBI observation 
for the jet base of M87 
(\citealt{2016ApJ...817..131H}, see also \citealt{1997ApJ...490..653H}).



Following previous works \citep{2000ApJ...528L..13F,2006ApJ...636L.109B,2016ApJ...831....4P},
we apply a simple Keplerian shell model for the accretion flow in M87.
As the purpose of this work is to newly clarify the effects being brought by 
an SSA-thick region in M87's BH shadow, we do not include the jet emission for simplicity.
Expected influences by the jet emission will be briefly discussed in \S 6.
In this Keplerian shell model,
the number density of thermal electrons and the temperature of the thermal electrons 
are described in the form of power-law radial distribution with scale height:
$n_{\rm e} = n_{\rm e}^0 (r/r_{\rm g})^{-1.5} \exp (-z^2/2H^2)$ and 
$T_{\rm e} = T_{\rm e}^0 (r/r_{\rm g})^{-1.0}$. 
We set 
$n_{\rm e}^0 = 4 \times 10^{6} ~{\rm cm^{-3}}$, 
$8 \times 10^{6} ~{\rm cm^{-3}}$, and
$1.6 \times 10^{7} ~{\rm cm^{-3}}$  
for $a_{*} = 0.998, 0.75$ and 0.5, respectively,  
and $T_{\rm e}^0 = 7{\times}10^{10}$ K for $a_* = 0.998$, while $8{\times}10^{10}$ K for $a_* = 0.5$, and 0.75,
so that the accretion flow can be partially SSA thick at 230~GHz. 
The magnetic field strength in the shell is given by 
$B^2/8\pi = \beta^{-1}n_{\rm e} m_{\rm p} c^2 r_{\rm g}/6r$
, where $\beta$ represents the magnetization of the accretion flow, i.e., the accretion flow is more magnetized when $\beta$ is lower. 
We set the inner edge of the accretion flow at $r = r_{\rm ISCO}$ 
and no emitting plasma exists inside $r_{\rm ISCO}$.
It is important to note that \cite{2002ApJ...573..754K} pointed out the
emission from the plunging region between the event
horizon and the innermost stable circular orbit (ISCO) of matter is not negligible in radiatively inefficient accretion flows \cite[see also][]{2012MNRAS.426.3241N}. 
In this paper, we will demonstrate BH shadow images with disk truncation at  $r = r_{\rm ISCO}$. 
\footnote{We have also calculated the BH shadow images without matter 
truncation and 
confirmed the same conclusions.}

In the accretion flow with a moderately high mass accretion rate in M87, 
the synchrotron cooling timescale should be shorter than the accretion 
timescale 
\citep{2012MNRAS.421.1517D,2016A&A...586A..38M}.
Nonnegligible energy exchange between protons and electrons 
can work via the Coulomb collisions 
\citep{2017ApJ...844L..24R} or via the plasma instability whose relaxation timescale can be faster than that of the Coulomb collisions
 \citep{1988ApJ...332..872B}.
In such situations, the cooled accretion flow can contract in the vertical direction with the magnetic field being trapped inside the disk, i.e., 
moderately geometrically thin, low beta accretion flow 
will be formed \citep{2006PASJ...58..193M}.
The estimated field strength of M87 using 
early EHT data indeed suggests $B\approx 50-120$~G \citep{2015ApJ...803...30K}, 
which seems to indicate a fast synchrotron cooling.
We, therefore, assume that the moderately geometrically thin, magnetically dominated accretion flow surrounds the supermassive black hole (SMBH) in M87. 
Hereafter, the scale height of the accretion flow is assumed to be $H = 0.1 R$,
where $R$ is the cylindrical radius, and we set $\beta = 0.3$.




\section{Dark crescent in the BH shadow}

Figure \ref{fig:shadow_spin} shows the resultant intensity image for the model with $a_* = 0.998$ (left), $a_* = 0.75$ (center), and $a_* = 0.5$ (right), and with viewing angle
$i = 30^{\circ}$ (top) and $15^{\circ}$ (bottom).
Importantly, the dark-crescent feature is newly found 
between the SSA-thick ISCO ring and the photon ring for highly spinning BHs ($a_{*} = 0.998$ and 0.75).
Below, we explain how the dark crescent is formed around the rapidly spinning BHs.

The key point here is an offset of the center of the SSA-thick ISCO ring and the photon ring.
As is well known, due to the frame-dragging effect of the Kerr BH, the center of the photon ring shifts in the direction ${\bm e}_{\rm spin} \times {\bm e}_{\bm k}$ in the observer's screen, where  ${\bm e}_{\rm spin}$ and ${\bm e}_{\bm k}$ are the unit vector of the BH spin and that of photons perpendicularly crossing the observer's screen, respectively. That is, the photon-ring center shifts in the positive direction in the $x$--$y$ plane of the observer screen (i.e., horizontal direction) in this study.
On the other hand, the SSA-thick ISCO ring does not show
 significant horizontal deviation.

Figure \ref{fig:trajectory} explains why this difference appears.
In this figure, the trajectory of photons forming the photon ring and the SSA-thick ISCO ring in the vicinity of the Kerr BH with $a_* = 0.998$ is projected onto the equatorial plane of the Boyer--Lindquist coordinate. 
The photon ring is the gravitationally lensed image of the (unstable) photon orbit.
Because the absolute values of angular momenta of the co- and counter-rotating photons on the photon orbit around highly spinning BHs 
(i.e., blue and red curves in Figure \ref{fig:trajectory}(a)) are distinctly different,  
the impact parameter is significantly different between them.
This results in the positional shift of the photon-ring center in the observer screen \citep{1973blho.conf..241B}.
On the other hand, 
photons emitted by the innermost accretion flow produce an SSA-thick ISCO ring
 and the ISCO ring does not show such a significant positional shift of its center (green and purple curves in Figure \ref{fig:trajectory}(b)).
This is because these rays deviate from the photon orbit near the BH and the resultant impact parameter does not strongly depend on the sign of angular momentum of photons.
Then, the positional offset between the center of the photon ring and that of the SSA-thick innermost region forms the "dark crescent" in the resultant image.
The dark crescent appears in the opposite side of relativistically beamed bright region.

Because the horizontal shift of the photon-ring increases with BH spin (as well as the ISCO radius for matter decreasing with BH spin), the larger and clearer dark crescent appears when the BH spin is higher:
the dark crescent appears in the models with $a_* = 0.998$ and 0.75, while it is not found around the BH with $a_* = 0.5$ in our shadow images.
We can find that a clearer dark crescent for the model with $i = 30^{\circ}$ than for $i=15^{\circ}$, 
because of more prominent horizontal shift of the photon ring by the 
frame-dragging effect in the case with larger viewing angle.

\begin{figure}[h!]
\begin{center}
\includegraphics[scale=0.44]{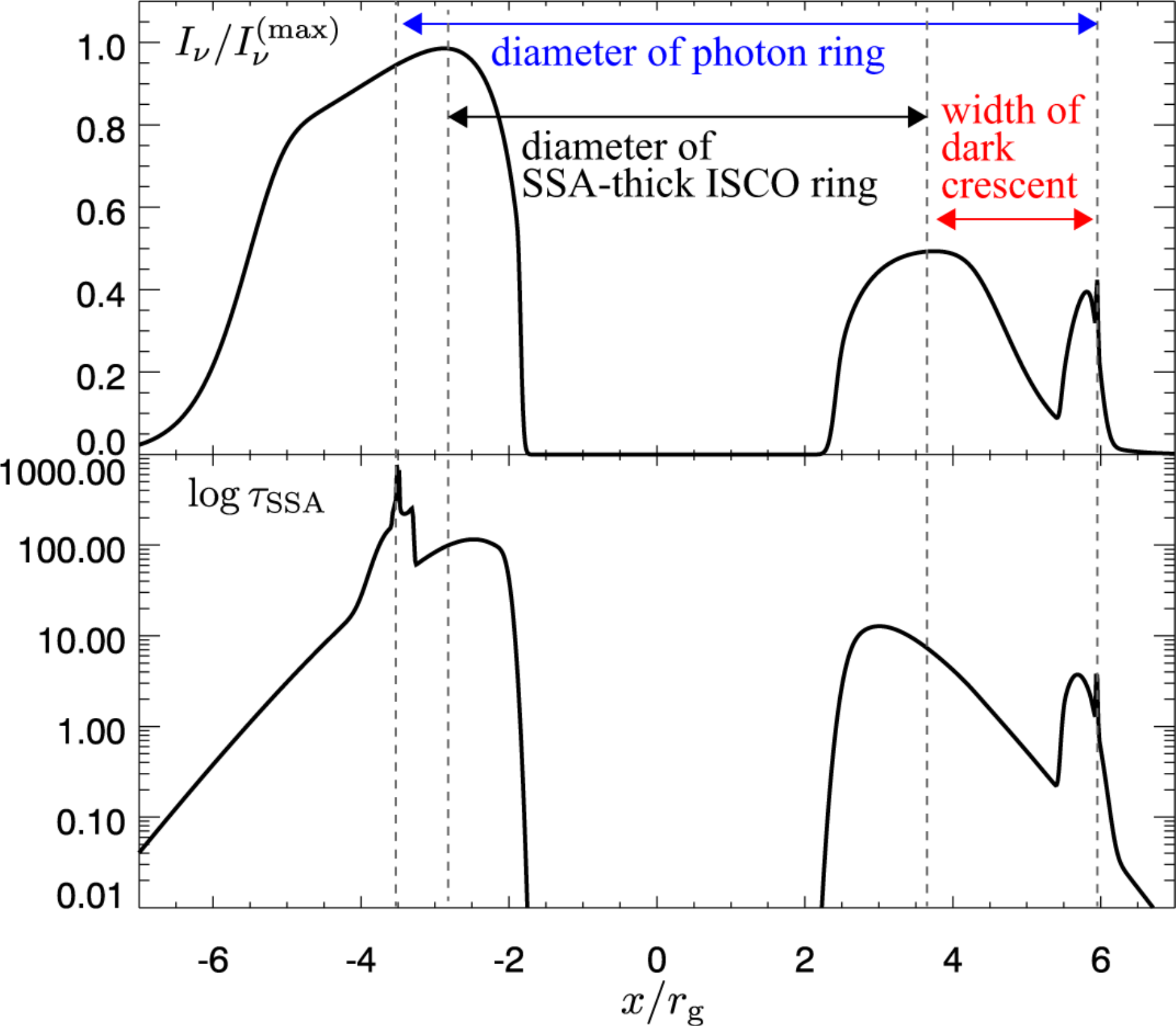}
\caption{Linear-scale intensity and log-scale ${\tau}_{\rm SSA}$ on $y=0$ in the observer's screen at $\nu = 230$ GHz.
We define the width of the dark crescent as the length between the SSA-thick ISCO ring and the photon ring on the line $y=0$ in the region $x>0$ in the observer screen.
\label{fig:I_tau_1D}
} 
\end{center}
\end{figure}



In Figure \ref{fig:shadow_tau}, we present the intensity images (top) and $\tau_{\rm SSA}$ (bottom) in the observer's screen at 230~and 350~GHz for $i=30^{\circ}$ and $15^{\circ}$ 
in the model with $a_* = 0.998$.
This is the first example explicitly displaying $\tau_{\rm SSA}$ in the observer's screen.
These spatial distribution maps of $\tau_{\rm SSA}$ on the observer screen enable us to understand
the formation of the dark crescent in the BH shadow image.
By comparing the intensity and $\tau_{\rm SSA}$ map at 230~GHz, we can confirm that the bright inner ring is formed via the emission from the SSA-thick innermost accretion flow. 
We note that the photon ring is $\tau_{\rm SSA} \gtrsim 1$, because the 
rays pass close to the photon orbit, 
which results in the large emission gain and the formation of the bright photon ring. 
At $\nu = 350$~GHz, the right side of both of the photon ring and ISCO ring becomes less luminous than those at 230~GHz because of their 
surface brightness at 350~GHz.
This makes the wider surface area of an SSA-thin region at 350~GHz, and the feature of the dark crescent becomes clearer than that at 230~GHz.


In order to define the width of the dark crescent, in Figure \ref{fig:I_tau_1D},
we present the intensity and $\tau_{\rm SSA}$ profile 
along $y=0$ on the observer's screen in the model with $i = 30^{\circ}$ and $a_* = 0.998$.
We normalized the intensity by its maximum value on the screen.
In the numerical calculations, we can well identify the photon-rings 
by the ${\tau_{\rm SSA}}$ 
profile because the photons forming the photon ring are trapped
 near the circular photon orbit and ${\tau}_{\rm SSA}$ becomes significantly 
large.
We define the position of the SSA-thick ISCO-ring as the position where the intensity is maximal in the inner bright region on $y=0$.
Subsequently, we define the width of the dark-crescent as the length between the SSA-thick ISCO ring and the photon ring on $y=0$ in the region $x>0$.

\begin{figure}[h!]
\begin{center}
\includegraphics[scale=0.55]{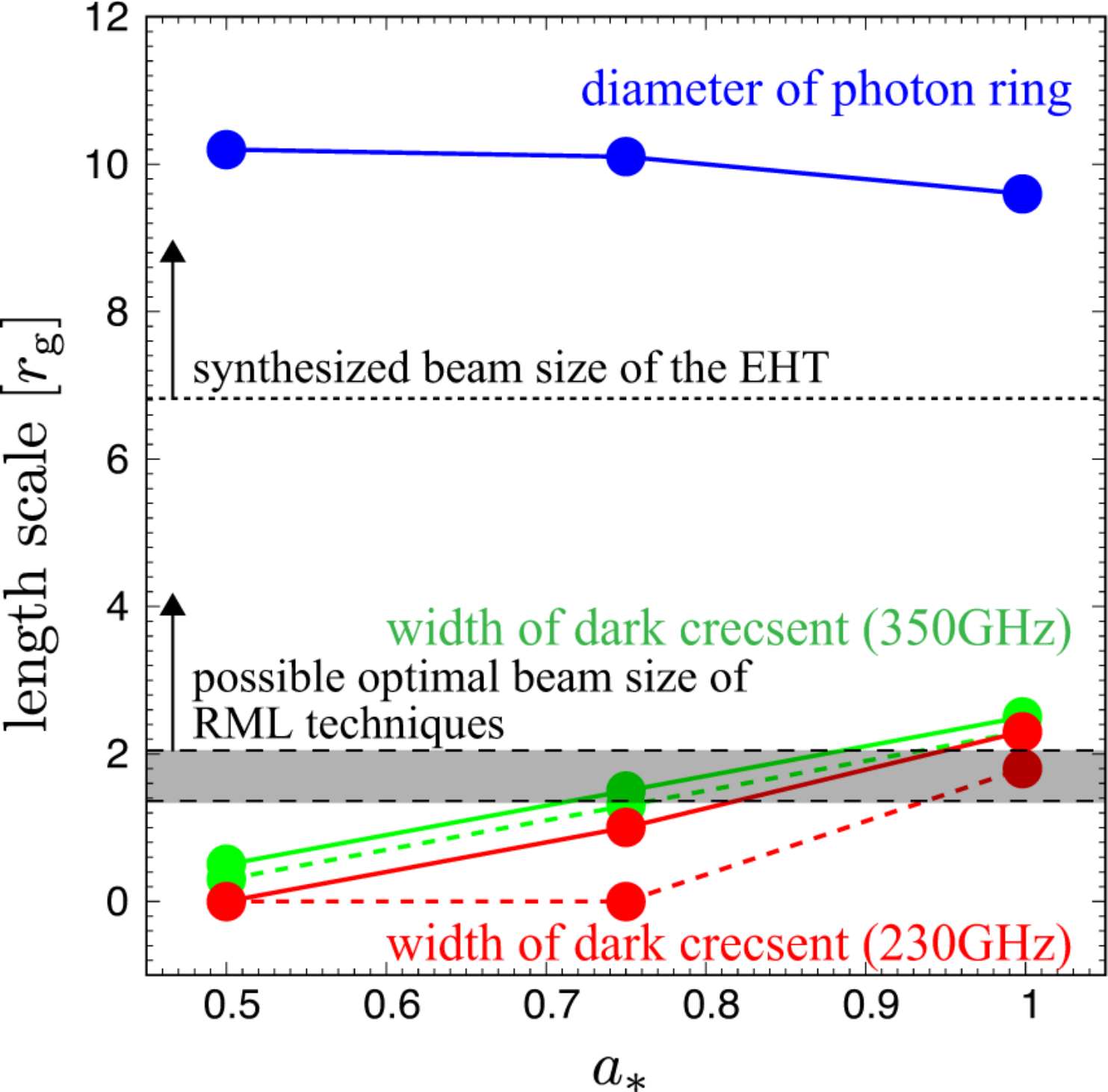}
\caption{Apparent width of dark crescent at 230/350~GHz (red/green) and photon ring (blue) for M87 with viewing angle $30^{\circ}$ (solid line) and $15^{\circ}$ (dashed line). The beam sizes of EHT with/without the sparse modeling for M87 are represented by the dashed/dotted black lines. Here, the expected range of potential  resolution-limit with RML imaging techniques 
\citep[${\simeq}$ 20--30 \% of the nominal synthesized beam of the EHT 2017 array; see][]{2016ApJ...829...11C,2017ApJ...838....1A,2017AJ....153..159A} is represented by the shaded region. 
}
\label{fig:shadow_size}
\end{center}
\end{figure}

\begin{figure*}[t!]
\begin{center}
\includegraphics[width=0.95\textwidth]{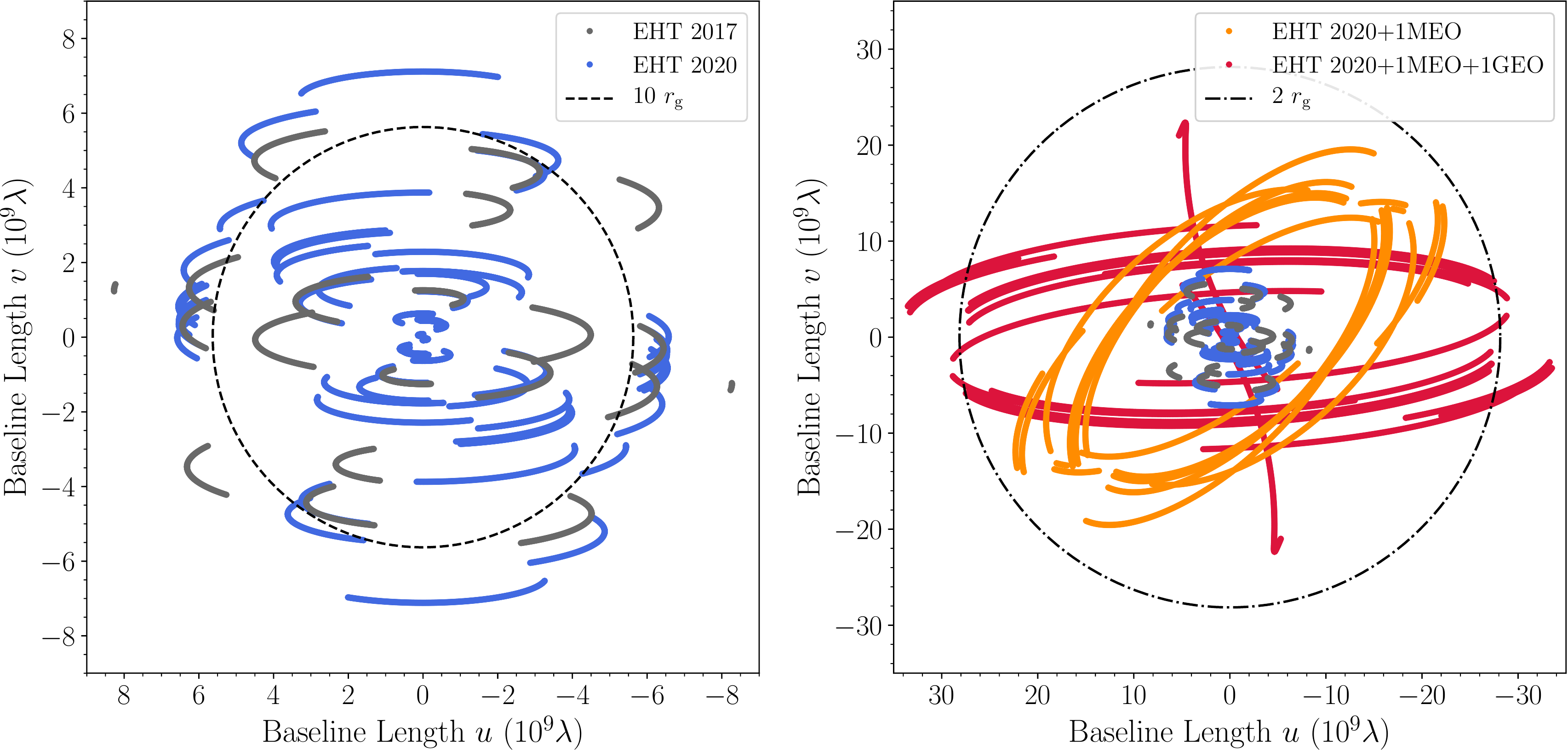}
\caption{
$uv$-coverages of observational simulations at 230~GHz (see \S \ref{DetectM87_2} for details about involved stations). (Left) $uv$-coverages of ground EHT stations. Black lines are from baselines including stations at five geographic sites used in EHT 2017 observations, while blue lines are additional baselines involving three sites expected to participate after the 2020 observations. The dashed line shows the fringe spacing for $10r_g$. (Right) $uv$-coverages of the EHT full array with additional space stations at MEO (orange) and GEO (red) orbits. Blue and black lines are the same as in the left panel. The dashed line shows the fringe spacing for $2r_{\rm g}$.
\label{fig:uv-coverage}
}
\end{center}
\end{figure*}

\begin{figure*}[t!]
\begin{center}
\includegraphics[scale=0.75]{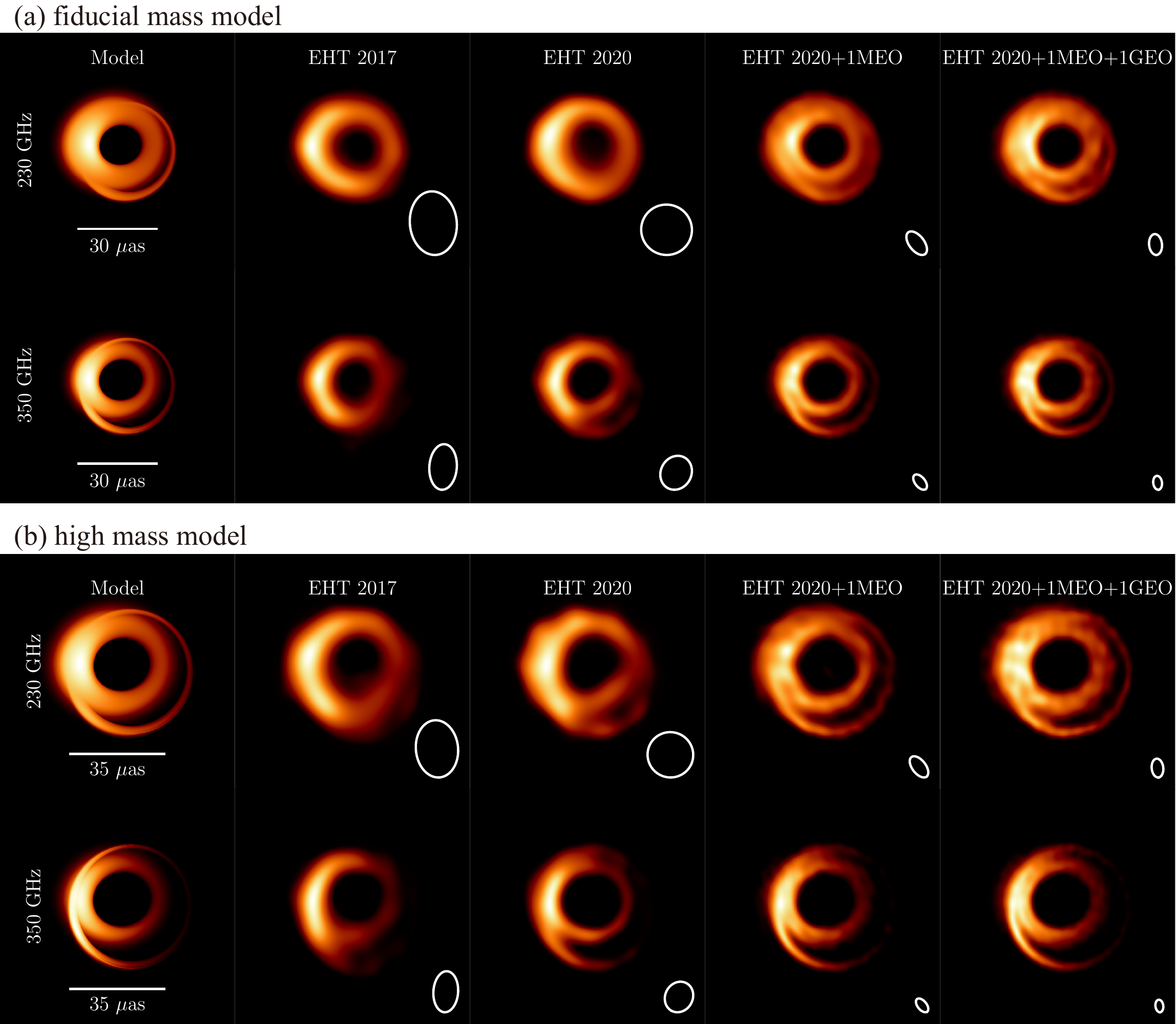}
\caption{Reconstructions of the simulated images of (a) fiducial-mass model $M = 6.2 \times 10^9 M_{\odot}$ and (b) high-mass model $M = 9 \times 10^{9}M_{\odot}$ for four array configurations shown in Figure \ref{fig:uv-coverage} (see \S \ref{DetectM87_2} for details). 
Color scales are linear. The white ellipse in each panel shows the FWHM size of the uniform-weighted synthesized beam. All images are not convoluted with the beam.}
\label{fig:reconstruction}
\end{center}
\end{figure*}

One may notice that connecting line between the photon-ring center and SSA-thick center does not seem to be perfectly perpendicular to the BH-spin vector, i.e., not parallel to the $x$-axis on the observer screen (see, e.g., Figure \ref{fig:shadow_tau}). The angle between this connecting line and the $x$-axis depends on the viewing angle. If we increase the viewing angle from the pole-on to edge-on, then the ISCO ring moved from the center to the top of the screen. Therefore, the dark crescent will appear in the bottom right region inside the photon ring when the viewing angle is higher (e.g., $45^{\circ}$ or $60^{\circ}$), while the dark crescent appears in the middle right region inside the photon ring in this work (i.e., $15^{\circ}$ or $30^{\circ}$). And the angle weakly depends on the photon frequency at least between 230 and 350 GHz, as the shape of the SSA-thick ISCO ring is weakly warped when the photon frequency changes.
At least, in this work, the deviation of the connecting line from the $x$-axis is small and the definition of the width of the dark crescent in Figure \ref{fig:I_tau_1D} is reasonable.

Last, it is worth discussisng why the dark crescent in M87 BH shadow 
is not clearly recognized in previous works 
\citep[e.g.,][]{2000ApJ...528L..13F,2004ApJ...611..996T,2009ApJ...697.1164B,2012MNRAS.421.1517D,
2016A&A...586A..38M}. 
Prior studies focused on the detectability of the  photon ring in the BH 
shadow, 
so that the parameters seems to be biased to the SSA-thin limit cases 
(or the SSA-thick limit with disk emission for comparison).
The existence of partially SSA-thick plasma in the BH shadow of M87
has been suggested just recently \citep{2015ApJ...803...30K}; 
it may be one  of the reasons why the dark crescent in M87 BH shadow 
was not explicitly recognized in previous works.
Now, if we carefully revisit the BH shadow images in previous works,
some features similar to our BH shadow images can be seen, for example, 
in Figure A.1 in \cite{2016A&A...586A..38M} for their models RH20, RH40, and RH100. 
Although no explicit statement is seen in the literature,
the BH shadow images shown in their Figure 3 (their models RH20, RH40, and RH100), 
seems to have similar properties to our BH shadow images, 
which might be formed via the SSA-thick jet base and the photon ring.

\section{Detectability of the dark crescent in M87}  \label{DetectM87}

It is important to discuss detectability of the dark crescent in M87.
Here we will argue (i) spatial resolution and sensitivity from the theoretical point of view, 
(ii) practical reconstruction tests of the theoretical images assuming current and future EHT arrays, and 
(iii) a possible contamination by the jet or blob emission.

\subsection{Spatial resolution and sensitivity from the theoretical point of view}  \label{DetectM87_1}
First of all, in Figure \ref{fig:shadow_size}, 
we summarize the dependence of the dark crescent's width (see Figure \ref{fig:I_tau_1D} for definition)
on BH spin.
It is well known that a diameter of the photon ring only has a very weak dependence on a BH-spin value.
On the other hand, we find that the dark-crescent width has a clear dependence on the BH-spin value.
The width of the dark crescent increases with BH spin: the width is $\sim ~ 2r_{\rm g}$ at $a_* = 0.998$, meanwhile the dark-crescent disappears at $a_* = 0.5$.
We overplotted the beam size of EHT ($\simeq 25 \mu{\rm as}$), 
which corresponds to $\sim 7 r_{\rm g}$ assuming the
 BH mass to be $6.2 {\times} 10^9 M_{\odot}$ and distance to be 16.7 Mpc.
Apparently, the spatial resolution of EHT is larger than the width of the dark crescent.
However, recent active developments of regularized maximum likelihood \citep[RML;][]{2019ApJ...875L...4E} imaging techniques such as sparse modeling and the maximum-likelihood method significantly improve image fidelity at modest super-resolutions finer than the diffraction limit of the interferometric array \citep[e.g.][]{2014PASJ...66...95H,2016ApJ...829...11C,2018ApJ...857...23C,2017ApJ...838....1A,2017AJ....153..159A}. 
For instance, \cite{2016ApJ...829...11C} and \cite{2017ApJ...838....1A,2017AJ....153..159A} presented that the effective spatial resolution may reach to 
${\simeq} 20 - 30$ \% of the diffraction limit with RML imaging techniques, of which corresponding resolution  is also shown in Figure \ref{fig:shadow_size}.
In Figure \ref{fig:shadow_size},
it is found that the dark crescent can be marginally resolved by EHT resolution
with the aid of the sparse modeling technique when the BH-spin value is $a_{*} {\gtrsim} 0.8$.

We note, in addition, that the predicted BH shadow images of M87 in this work have enough flux densities
as they have been already matched up to the result of early EHT observation \citep{2012Sci...338..355D, 2015ApJ...807..150A}.
In our model, the magnetic field strength and the electron temperature  at $r = r_{\rm ISCO}$ are  
$ 90~{\rm G} \lesssim  B \lesssim 200~{\rm G}$ and 
$ 1.9\times 10^{10}~{\rm K} \lesssim  T_{\rm e} \lesssim 5.7\times 10^{10}~{\rm K}$, respectively.
These are well consistent with the estimate of  \cite{2015ApJ...803...30K} for M87 at 230~GHz. 
The corresponding synchrotron flux in all our models is $\sim$ 0.5-0.8 Jy at 230~GHz, 
which is $\sim$ 50-80 \% of the observed flux (${\sim}$ 1 Jy) at the same photon frequency in M87 \citep{2012Sci...338..355D}.
If the emission from the jet is taken into account, the observed flux may be slightly higher than 1 Jy, i.e., our disk model will likely appear in the flaring state in M87.
We emphasize that the radio fluxes of our models are, indeed, greater than the observed flux during the  EHT 2017 observations {$\sim$ 0.5} Jy \citep{2019ApJ...875L...1E, 2019ApJ...875L...2E, 2019ApJ...875L...3E, 2019ApJ...875L...4E, 2019ApJ...875L...5E, 2019ApJ...875L...6E}.

We also note that the estimated mass accretion rate on the  ISCO scale 
is $\sim 10^{-3} M_{\odot} ~ {\rm yr}^{-1}$ in our flaring state model. 
This accretion rate would be higher than the mass accretion rate in the quiescent state in M87.
This is because the mass accretion rate at $\sim 40 r_{\rm g}$ in quiescent state in M87 is estimated to be $\sim 9 \times 10^{-4}M_{\odot}~{\rm yr}^{-1}$ by using the measurement of the rotation measure \citep{2014ApJ...783L..33K}, and the mass accretion rate on the ISCO scale is smaller than that at $\sim 40 r_{\rm g}$  due to a wind mass-loss process.
Therefore, the mass accretion rate in our flaring state model would be indeed higher than that of the quiescent state in M87, i.e., the dark crescent appears in flaring states with a relatively high mass accretion rate.

\subsection{Practical reconstruction tests of theoretical images assuming current and future EHT arrays}  \label{DetectM87_2}

For a more precise discussion on the detectability of the dark crescent in M87, we perform observational simulations using theoretical BH shadow images. Synthetic observational data are created with the \texttt{eht-imaging} library\footnote{\url{https://github.com/achael/eht-imaging}} \citep{2016ApJ...829...11C,2018ApJ...857...23C} and imaged with \texttt{SMILI}\footnote{\url{https://github.com/astrosmili/smili}} \citep{2017ApJ...838....1A,2017AJ....153..159A}.
We perform synthetic observations with the following four array configurations\footnote{see \url{https://eventhorizontelescope.org/array} for the abbreviation of each telescope}: (1) EHT 2017 array consisting of seven stations at five geographic sites, ALMA and APEX in Chile, LMT in Mexico, SMT at Mt. Graham in Arizona, IRAM 30m Telescope at Pico Veleta in Spain, SMA and JCMT at Maunakea in Hawaii; (2) an extended ground-based array (henceforth EHT 2020) expected for 2020s with the following four additional stations, ARO 12m Telescope at Kitt Peak in Arizona, NOEMA at Plateau de Bure in France, GLT in Greenland, and a single-dish telescope at Owens Valley in California; (3) EHT 2020 array with a space station at the middle Earth orbit (MEO); (4) EHT 2020 array with two space stations at MEO and geosynchronous orbit (GEO). We use the orbits of the {\it Galileo}-IOV PFM and {\it TerreStar-1} satellites for the stations on MEO and GEO orbits on 2019 January 1, respectively. Figure \ref{fig:uv-coverage} shows $uv$-coverages of simulated M87 observations at 230~GHz with the above four array configurations. Synthetic data are generated for the full single track at the bandwidth of 2~GHz. Images are reconstructed from synthetic complex visibility data sets with $\ell_1+$TSV regularization of sparse modeling \citep{2018ApJ...858...56K}.

Figure \ref{fig:reconstruction} demonstrates the expected images in current and future EHT observations. 
In panel (a), the reconstructed images for the fiducial-mass model ($M = 6.2 \times 10^9 M_{\odot}$) at 230~GHz (upper images) and 350~GHz (lower images) are presented.
For ground arrays, the signature of the dark crescent starts to appear at 350 GHz with the EHT 2020 array. One can find that the future EHT arrays with additional space stations will significantly improve the angular resolution less than 2$r_{\rm g}$ (Figure \ref{fig:uv-coverage}), which is sufficient to detect the dark crescent very clearly at 230~GHz and finer structures on scales of a few $r_{\rm g}$ (see also \citealt{Fish_2019} for similar space-VLBI simulations with MEO/GEO satellites). Importantly, at 350~GHz, the dark-crescent feature appears more clearly, because the opacity for SSA decreases with photon frequency and the dark-crescent region becomes larger (see also Figure \ref{fig:shadow_tau} and \ref{fig:shadow_size}).

In panel (b), we explore the detectability of the dark crescent in a possible case with extremely high BH mass in M87 ($M = 9 \times 10^9 M_{\odot}$). 
We examine this extreme case because the the diameter of the SSA-thick ISCO ring would explain the expected diameter of shadow $\sim 40 \mu {\rm as}$ in the early EHT observations \citep{2012Sci...338..355D,2015ApJ...807..150A}. 
We calculate the BH shadow image of this extremely high mass BH with $a_{*} = 0.998$ by setting the same parameter as the fiducial-mass model, except the lower electron number density $n_{\rm e}^0 = 2\times10^6~{\rm cm}^{-3}$.
At first, we can find that the diameter of the SSA-thick ISCO-ring is $\sim 40 ~\mu {\rm as}$ in this case, because $M/D$ is larger than the fiducial mass model. Interestingly, without the space-VLBI arrays, the dark-crescent feature very faintly but certainly appears at 350~GHz (the second left and the middle columns in Figure \ref{fig:reconstruction}(b)). The dark crescent very clearly appears when we use the EHT with space VLBI both at 230 and 350~GHz.

Our synthetic observations suggest that, with the ground-based arrays only exisiting with existing ground millimeter telescopes, the detection of dark-crescent features would be challenging at least at 230~GHz contrary to the expectation in \S \ref{DetectM87_1}. It is most likely due to their sparse baseline coverages, which make it difficult that the signature of the faint dark crescent is significantly detected in observed visibilities --- equivalently, on the image domain, the image cannot get sufficient dynamic ranges and effective angular resolutions to detect the faint feature. Recently, \cite{Doeleman_2019} and \cite{Palumbo_2019} have presented potential extensions of the EHT to fill $uv$-coverages within the Earth diameter by adding many new ground stations with small dishes to the EHT and/or low-Earth-orbit (LEO) space stations, which provide much denser $uv$-coverage and consequently drastic improvements in the dynamic range of images. The future addition of many new ground telescopes and/or LEO space stations could allow the detection of the dark-crescent feature only with a (nearly) Earth-sized array.

We emphasize that the observation both at 230 and 350~GHz is very important to explore the dark-crescent feature in BH shadows.
As we mentioned above, the width of the dark crescent is larger at 350~GHz than at 230~GHz.
On the other hand, the observation of the photon ring would be better at 230~GHz, because the retrograde-orbit side of the photon ring would clearly appear at 230~GHz rather than at 350~GHz in some cases (e.g., see Figure \ref{fig:reconstruction}(b)). 
It is, therefore, important to measure the width of the dark crescent with the simultaneous use of reconstruction images both at 230 and 350~GHz.

\subsection{A possible contamination by the jet or blob emission}  \label{DetectM87_3}

In the case of M87, the jet emission might contaminate
the dark-crescent feature.
However,
according to previous works on M87 BH shadow images
\citep{2009ApJ...697.1164B,2016A&A...586A..38M},
the jet base emission placed at the brighter side of the photon ring,
i.e., at the opposite side of the dark crescent.
If this is the case for M87, 
then the jet emission does not contaminate the dark-crescent feature.
The physical mechanism of plasma supply to the magnetic-funnel jet region
is highly uncertain 
\citep{2011ApJ...730..123L, 2011ApJ...735....9M,2012ApJ...754..148T,2015ApJ...809...97B,2016ApJ...818...50H, 2017ApJ...845..160P, 2018A&A...616A.184L, 2018ApJ...863L..31C,2019PhRvL.122c5101P}, and 
further intensive investigations, therefore, definitely should be needed for the jet emission model. 
These are planned for future work.

One may note that, in addition, a blob in the accretion flow or in the jet would contaminate the clean signature of the dark crescent in snapshot images.
We expect that the long-term observation will clean such a contamination, because the contamination due to the blobs will be smoothed out in the time-averaged data.
The cleaner dark-crescent image would be obtained if we average the images over the rotation timescale of blobs in the vicinity of the BHs (i.e., more than several days in M87).


\section{Summary}

In this work,
we carried out GR ray-tracing radiative transfer calculations in the Kerr spacetime, and 
we have investigated the BH shadow images of M87
properly taking into account the SSA-thick ISCO ring.
The summary is shown as follows.

\begin{itemize}
 \item 
 We have discovered a new feature in BH shadow, 
when a rapidly spinning BH is surrounded by partially SSA-thick plasma.
In this case, the positional offset between the center of the photon ring 
and the SSA-thick ring at the ISCO due to the frame-dragging effect in the Kerr spacetime.
As a result, a dark-crescent structure is generally 
produced between the photon ring and the SSA-thick ISCO ring in the BH shadow image.
This dark crescent is a new manifestation of the high spin of BHs.

\item 
Actual detectability of the dark crescent in M87 is of great interest. 
As shown in Figure \ref{fig:reconstruction}, we have found that the dark crescent can be resolved by the future full EHT array with space stations at 230 and 350 GHz. 
We expect that the dark crescent will likely appear in the flaring state in M87, because the mass accretion rate is bit higher than that estimated by the rotation-measure observation in the quiescent state \citep{2014ApJ...783L..33K}, and the radio fluxes of our models (0.5-0.8Jy) with the accretion flow only at 230 GHz is greater than the observed fluxes during the EHT 2017 observations (0.5Jy) reported in \cite{2019ApJ...875L...1E,2019ApJ...875L...2E,2019ApJ...875L...3E,2019ApJ...875L...4E,2019ApJ...875L...5E,2019ApJ...875L...6E}. This is consistent with the fact that the diameter of the  SSA-thick ISCO ring surrounding the fiducial-mass BH is smaller than the estimated shadow size in the early EHT observation \citep{2012Sci...338..355D}. 

\item
We also note that there is another possibility that the mass of the SMBH is slightly higher ($M = 9 \times 10^{9} M_{\odot}$). It has been shown that the diameter of the SSA-thick ISCO ring is ${\sim} 40 ~ \mu{\rm as}$ in this high BH mass model. The future EHT observation with space stations will identify these two models with different BH mass, because it can resolve the SSA-thick ISCO ring and the photon ring, and the diameter of the photon ring tells us the mass of the BH.

\item
The emission of a jet or blob in M87 may matter for detecting  the dark-crescent feature.
So far, according to  previous works on M87 BH shadow images
\citep{2009ApJ...697.1164B,2016A&A...586A..38M},
we can assume that the jet base emission appears at the brighter side of the photon ring,
i.e., at the opposite side of the dark crescent.
In such a case, the jet emission does not contaminate the dark crescent.
We will separately investigate jet emission in detail in the future.
With respect to the emission from the blob in the accretion flow or in the jet, the long-term observation will enable us to clean up the contamination in the images.
This is because the contamination due to the blobs will be smoothed out in the time-averaged data over the rotation timescale of blobs in the vicinity of the BHs (i.e., more than several days in M87).

\item 
We emphasize the importance of 350~GHz VLBI observations 
in the future, 
because the width of the dark crescent is larger at 350 GHz due to the decrease of the optical depth for SSA (Figure \ref{fig:shadow_tau} and \ref{fig:shadow_size}), which results in the more robust detectability of the dark crescent in future EHT observation (Figure \ref{fig:reconstruction}).
Toward realizing the transcontinental VLBI at 350~GHz, 
the 12 m diameter radio telescope is now almost deployed to the Summit Station in Greenland. 
The telescope (Greenland Telescope, GLT) is to become one of the VLBI
stations at 350~GHz, providing the longest baseline, longer than 
9000 km to achieve an exceptional angular resolution of 20 $\mu$as, 
and we will be able to conduct observations at event horizon angular resolution
\citep{2014RaSc...49..564I,2017arXiv170504776A}.


\end{itemize}


\acknowledgments 
  We thank K. Ohsuga, V.~L. Fish, H.~R. Takahashi, M. Nakamura, and K. Toma 
for useful comments and discussion. 
The numerical simulations were carried out on the XC30 and XC50 at the Center for
  Computational Astrophysics, National Astronomical Observatory of
   Japan.
This work was supported by JSPS KAKENHI grant Nos. JP18K13594 (T.K.), JP18K03656 and JP18H03721 (M.K.), and a grant from the National Science Foundation (AST-1614868; K.A.). 
This work was also supported in part by MEXT SPIRE, 
   MEXT as ``Priority Issue on post-K computer'' 
   (Elucidation of the Fundamental Laws and Evolution of the Universe), 
   ,JICFuS, and the NINS project of Formation of International Scientific Base and Network .
K.A. is a Jansky fellow of the National Radio Astronomy Observatory. The National Radio Astronomy Observatory is a facility of the National Science Foundation operated under cooperative agreement by Associated Universities, Inc. The Black Hole Initiative at Harvard University is financially supported by a grant from the John Templeton Foundation.

\software{
\texttt{RAIKOU} (Kawashima et al. in preparation), 
\texttt{SMILI} \citep{2017ApJ...838....1A,2017AJ....153..159A}, 
\texttt{eht-imaging} \citep{2016ApJ...829...11C,2018ApJ...857...23C}
}







\end{document}